\documentclass[12pt]{iopart}  
\pdfoutput=1

\usepackage{graphicx}
\usepackage{color}
\usepackage{ulem}
\usepackage{subfloat}
\usepackage{xr-hyper}
\usepackage{url}

\newcommand{\kB}{k_{B}}
\newcommand{\phiv}{\phi_{v}}

\newcommand{\squarebk}[1]{\left[#1\right]}

\definecolor{darkred}{rgb}{0.6,0,0}
\definecolor{darkgreen}{rgb}{0.0,0.4,0}

\begin{document}

\title[H.-Y. Deng et al. Analytical Theory for MSD in Glasses]{Configuration-tree Theoretical Calculation of the Mean-Squared Displacement of Particles in Glass Formers}

\author{Hai-Yao Deng$^1$\footnote{Equal contribution; Current address: School of Physics and Astronomy, University of Manchester, Manchester, M13 9PL, UK}, Chun-Shing Lee$^2$\footnote{Equal contribution}, Matteo Lulli$^2$, Ling-Han Zhang$^3$ and Chi-Hang Lam$^2$}
\address{$^1$ School of Physics, University of Exeter, Exeter EX4 4QL, United Kingdom \\ $^2$ Department of Applied Physics, Hong Kong Polytechnic University, Hong Kong \\ $^3$  Department of Physics, Carnegie Mellon University, Pittsburgh, Pennsylvania 15213, USA}

\vspace{10pt}

\begin{abstract}
We report an analytical evaluation of the mean-squared displacement (MSD) of the particles in glasses based on their coarse grained trajectories, after their vibrations are averaged out. The calculation is conducted by means of a local random configuration-tree theory that was recently proposed by one of us [C.-H. Lam, J. Stat. Mech. \textbf{2018}, 023301 (2018)]. It assumes that system dynamics is dominated by activated particle hops initiated by voids and that particles are distinguishable with, in general, diverse properties. Results are compared with the numerical simulations of a lattice glass model, and satisfactory agreement has been obtained for various density of particles over a wide range of temperatures in the entire region of time with only two parameters. 
\end{abstract}

\section{Introduction}
\label{sec:Introduction}
Upon rapid cooling, many liquids undergo a dramatic -- often more than ten decades -- increase in their viscosity and eventually become glasses at a temperature $T_g$. The question of how the constituent particles of a liquid lose their mobility in this process remains a source of dispute, having generated a library of theories of varying sophistication and complexity~\cite{berthier2011review,garrahan2011review,stillinger2013review}. 

Glasses are notoriously complex systems~\cite{angell2000jap}. Analytical yet accurate evaluations of any of their properties are very rare and valuable. The mean-squared displacement (MSD) of particles, defined as
\begin{equation}
	\label{MSD_def}
	\mbox{\textbf{MSD}}(t) = \mathcal{N}^{-1} \sum_j \mathbf{d}^2_j(t),
\end{equation}
where $\mathbf{d}_j(t)$ is the displacement of the $j$-th particle during a time $t$ and $\mathcal{N}$ is the total number of particles in a glass, is often computed to study glassy transition. As a ubiquitous signature of glass formation, upon approaching $T_g$, \textbf{MSD}$(t)$ develops a plateau, which is a precursor of the so-called $\alpha$ relaxation characterizing any glasses~\cite{Gotzebook}. 

In this contribution, we present an analytical calculation of \textbf{MSD}$(t)$ on the basis of a theory that was recently proposed by one of us~\cite{lam2018}, belonging to the class of dynamic facilitation descriptions \cite{fredrickson1984,palmer1984,ritort2003review,garrahan2011review,chandler2011,isobe2016,mandadapu2018}. Going beyond related calculations in Refs.~\cite{helfferich2014,phan2018}, our calculation covers the full range of time scales, spanning both the time window before and that after the plateau and hence providing a complete picture of the glassy dynamics. The theoretical result is then benchmarked against the `distinguishable particle lattice model' (DPLM)~\cite{zhang2017}, which is a lattice glass model realizing a dynamic facilitation mechanism motivated by glassy film simulation observations \cite{lam2018film}. Satisfactory agreement has been achieved --  including the position of the MSD plateau -- for various particle densities and over a wide range of temperatures by adjusting only two parameters of the model, which can further be fixed by the diffusion coefficient $D$ of the particles resulting in no additional free parameter.

\section{Theory}
\label{sec:Theory}
\subsection{The concept of a random local configuration tree} 
\label{sec:The_concept_of_a_random_local_configuration_tree}

Let us start with a brief recapitulation of the basic concepts of the theory that has already been described in details in Ref.~\cite{lam2018}. For the completeness of this paper, however, we have also provided in the Appendix a more comprehensive discourse of the main points and a comparison with the common approaches in the literature. 

At temperatures close to $T_g$, a particle in a liquid mostly vibrates about a temporary local equilibrium center, and occasionally makes a transient hop from one center to another. In the coarse-grained picture with the fast vibrations averaged out, only such hops are interesting and the phase space of the system reduces to the space of particle configurations. The system would then transit from one configuration to another via one or a sequence of particle hops -- which are assisted by the vibrations acting as an effective heat bath -- and relax its structure. The hopping distance is typically comparable to the particle size as measured in many works (see Ref.~\cite{lam2017} and references therein) from the van Hove correlation function exhibiting a peak. 
Each hop is primarily assumed to be initiated by a void. The void propagates along a string of particles that simultaneously displace one another during the propagation. These motions, called micro-strings,  generally come with different lengths obeying a distribution narrower than an exponential distribution \cite{glotzer2004}. As an approximation, we then assume that the micro-strings are of the same length and each possesses $\tilde{l}$ particles. In general, a void can propagate along one of $z$ strings lying in different directions. However, these strings each sense their own local energy landscape and they are not equally favorable. 

To simplify the analysis, let us divide the liquid into a number of local regions, each of volume $\mathcal{V}$. We assume that $\mathcal{V}$ is big enough so that a void propagation in one region does not affect the structural relaxation of neighboring regions, yet small enough so that one void propagation energetically affects the propagation of other voids in the same region (i.e. the fully interacting approximation~\cite{lam2018}). Under this approximation, it suffices to consider the dynamics of one local region. Let us represent the configurations of this region by the nodes of a graph and possible transitions by the edges. Any pair of nodes connected by one edge signify the presence of a micro-string. In a region containing $m$ voids, a node has $mz$ edges linked to it. The fully interacting approximation allows us to ignore the correlations between the edges. The region can revisit a previous node by retracing of a path of adjoining edges in reversed direction. However, we neglect the possibility of a loop of edges, so that the graph becomes a tree with its root corresponding to the initial configuration. The validity of this approximation stems from the very high dimensions of the particle configuration space and this has been discussed in details in Ref.~\cite{lam2018}. It should be further noted that, even with the inclusion of some loops, one may still represent the graph as a tree. This amounts to some additional correlations between edges, which should not alter our results significantly. 

It should be seen that the configuration tree is random in the sense that, in general its energetically favorable edges are randomly distributed. In this paper, we assume that the transition probabilities for the energetically favorable edges are finite and fixed but negligible for the unfavorable edges. In Figure~1, an illustration of the energy landscape is shown for the local configuration tree with one void and $z = 4$.

\begin{figure}[tb]
	\label{f:msd}
	\includegraphics[width=0.5\linewidth]{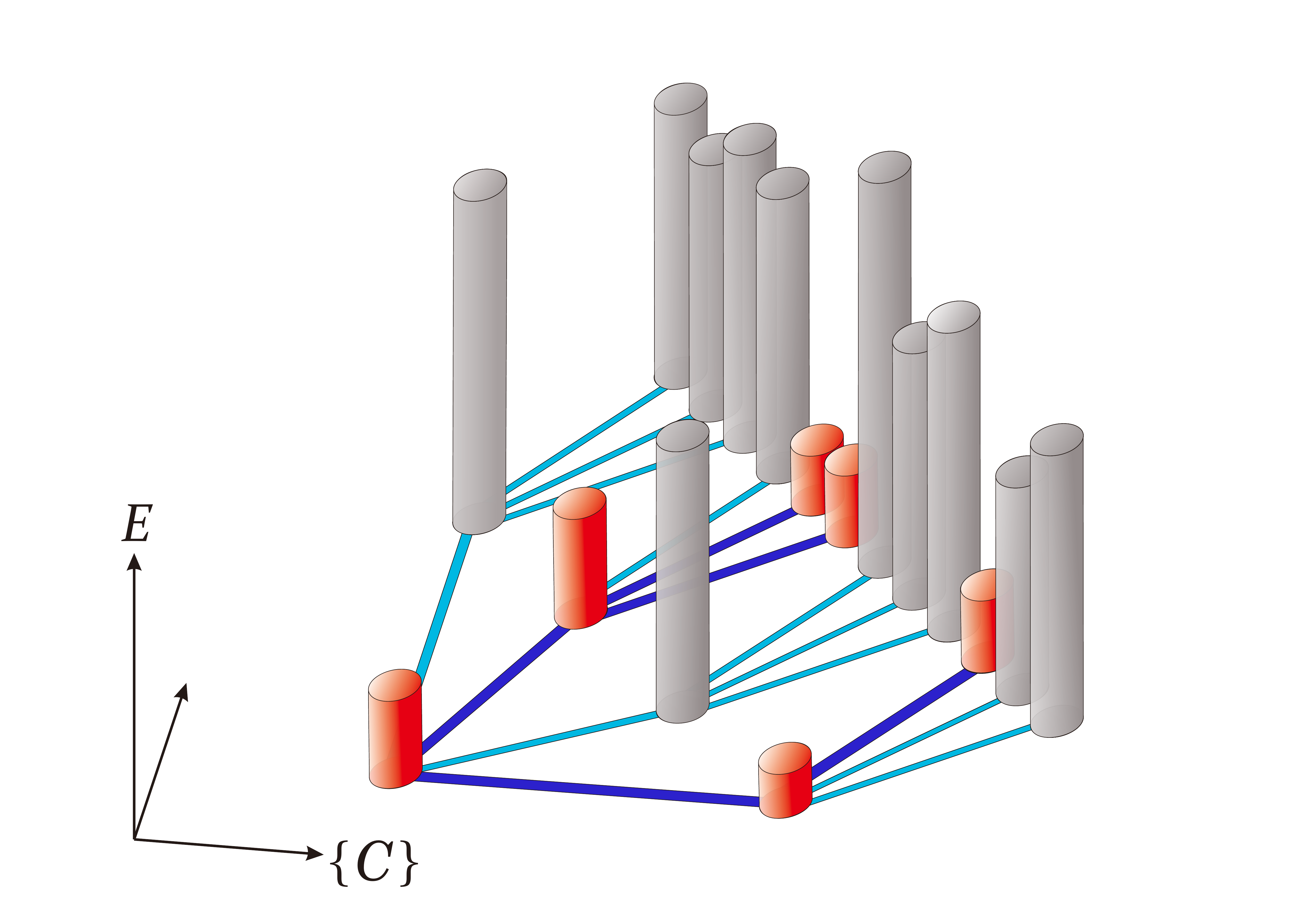}
	\includegraphics[width=0.5\linewidth]{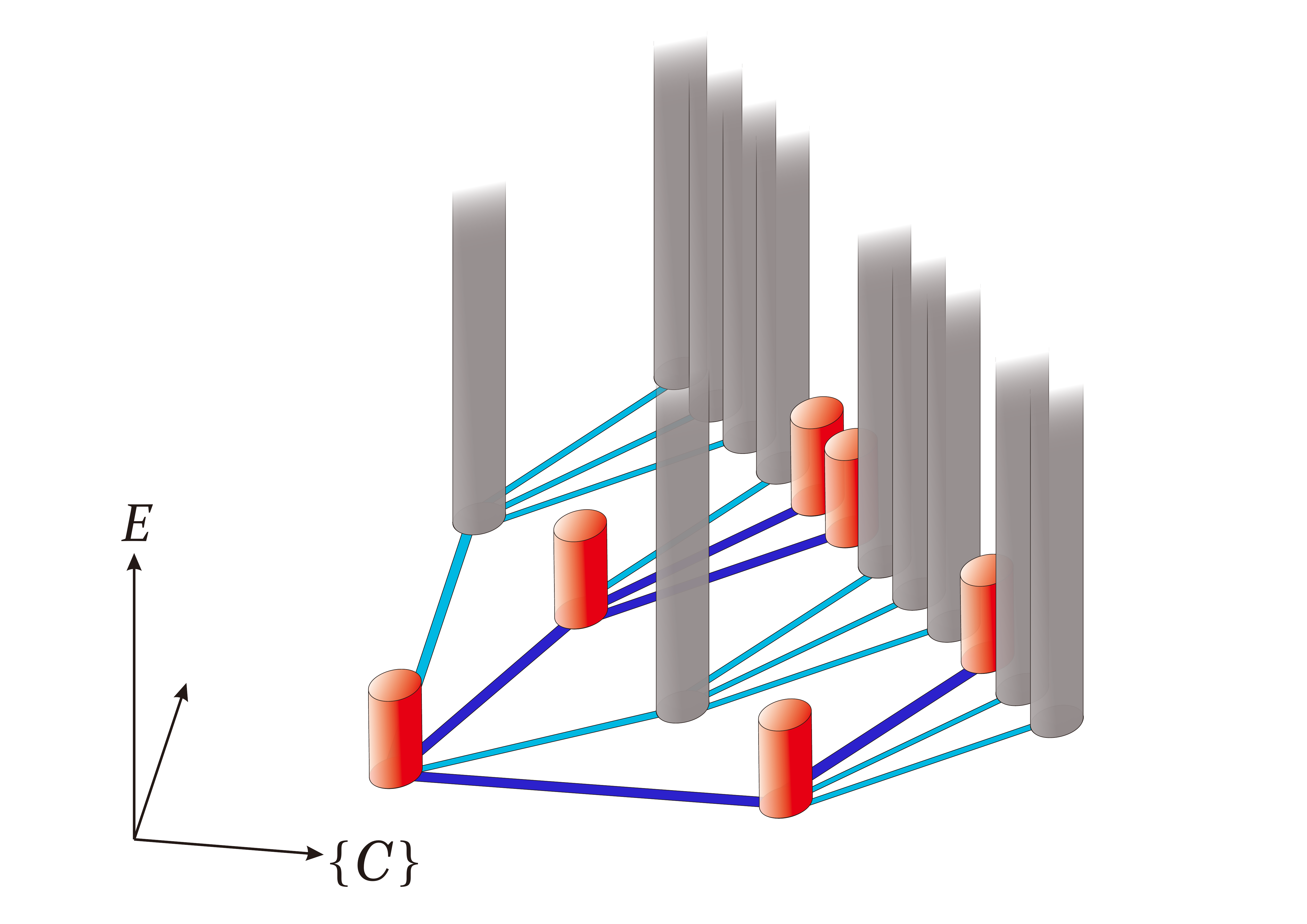}
	\caption{Sketch of the local configuration tree energy landscape. The energy for a node is represented by the height of the cylinder at the node. Left panel: red (grey) cylinders indicate the configurations with lower (higher) energies, which are considered to be more (less) energetically favorable. Blue (Cyan) edges are connecting the more (less) energetically favorable nodes with higher (lower) transition probabilities. Right panel: Red cylinders are assumed to have the same heights, approximating the fact that these configurations have almost the same low energy. Grey cylinders are now infinitely tall. Blue edges connect the nodes with red cylinders and represent the only possible transitions. Cyan edges are always connected to nodes with grey cylinders, for which transitions are impossible.}
\end{figure}

We should also mention that, the tree theory sketched above does not consider vibrations explicitly and more precisely captures the dynamics of the so-called inherent structures~\cite{stillinger1983}. An inherent structure is nothing but a microscopic configuration of meta-stable vibration centers. A void propagation then takes the system from one inherent structure to another. As to be discussed later, our theory is consistent with the numerical simulations carried out in Ref.~\cite{liao2001} of the evolution of the inherent structures of a binary Lennard-Jones liquid. 

In the rest of this section, we derive and solve a set of rate equations governing the motions of a local region as described by the tree. 

\subsection{Level dynamics on the tree: rate equations}
\label{sec:Dynamics_on_the_tree_rate_equations}
Let us consider a local region with $m$ voids and suppose the region is at equilibrium initially sitting at the root of its configuration tree defined to be level $n=0$.
In time $t$, the region will descend down the tree to some level $n$ with a probability denoted by $p_n(t;m)$.
Our purpose here is to derive the rate equations satisfied by this probability. 
Our formulation is analogous to previous calculations for random walks on regular trees \cite{cassi1989,hughes1982}.
It is emphasized again that, in our theory the random walk takes place primarily in the configuration space rather than in the real space. 

To ease the discussion, let us look at an ensemble of $N$ identical regions all initially at the root. At instant $t$, the number of members in this ensemble at the root ($n=0$) is $Np_0(t;m)$. After a short interval $dt$, this number becomes $Np_0(t+dt;m)$, the difference due to a loss of those members transiting to level $n=1$ and a gain of members at level $n=1$ transiting to the root. The transition from the root to level $n=1$ can take place through $mz$ edges~\cite{lam2018}, which we label by $\nu = 1,2,...,mz$. Let us denote by $dt \times \tilde{w}^{\nu}_{0\rightarrow 1}$ the probability of such a transition via edge $\nu$. The total transition rate is then 
$\sum_\nu\tilde{w}^\nu_{0\rightarrow1}$ which averages to $mz\tilde{w}_0$, with $\tilde{w}_0$ being the rate per edge. 
As the transitions are assisted by an effective heat bath of vibrations,  $\tilde{w}^\nu_{0\rightarrow1}$ in general depends on the energy difference $\Delta E$ between the configurations connected by the $\nu$-th edge. Denoting by $P(\Delta E)$ the probability for the occurrence of $\Delta E$ assuming an initial equilibrium configuration, a formally exact expression for the average transition rate $\tilde{w}_{0}$ along any of the edges from the root is
\begin{equation}
	\label{w_0}
	\tilde{w}_{0} = \int^{\infty}_{-\infty} d(\Delta E) w(\Delta E) P(\Delta E),
\end{equation}
where the rate $w(\Delta E)$ depends on the detailed microscopic dynamics.
Now the number of members lost to level $n=1$ is $Np_0(t;m) dt ~ mz\tilde{w}_0$.  

On the other hand, the transition for any member from level $n=1$ to the root can take place only through the edge linking the root and the node where the member is currently located. The probability of this transition for a member sitting on the node of the $\nu$-th edge is then $dt \times \tilde{w}^\nu_{1\rightarrow 0}$. Denote by $N^\nu_1$ the number of such members, which satisfies $\sum_\nu N^\nu_1 = Np_1(t;m)$. The number of members gained to the root is then $\sum_\nu N^\nu_1 dt ~ \tilde{w}^\nu_{1\rightarrow 0}$. At low temperatures, very few edges are energetically favorable and $N^\nu_1$ are significant only for the nodes linked to these edges. The rest can then be ignored. As $\Delta E$ for energetically favorable edges linking only low energy configurations are expected to be small, we can neglect the difference in their transition rates. As such, we shall replace $\tilde{w}^\nu_{1\rightarrow0}$ with an average rate $\tilde{w}$ for them. To actually compute $\tilde{w}$, let us define the energetically favorable edges as those for which $\Delta E \leq \mathcal{C}k_BT$, where $\mathcal{C}$ is a constant of order unity~\cite{lam2018}. Now, the rate $\tilde{w}$ for a favorable edge is approximated as follows
\begin{equation}
	\label{w}
	\tilde{w}^{-1} = \frac{\int^{\mathcal{C}\kB T}_{-\infty} d(\Delta E) w(\Delta E)^{-1} P(\Delta E)}{\int^{\mathcal{C}\kB T}_{-\infty} d(\Delta E) P(\Delta E)},
\end{equation}
which will be further explained below.
As such, the number of members gained amounts to $Np_1(t;m) dt ~ \tilde{w}$.

Combining these considerations, we obtain according to the detailed balance principle that $$Np_0(t+dt;m) - Np_0(t;m) = -Np_0(t;m)dt mz\tilde{w}_{0} + Np_1(t;m) dt \tilde{w},$$ or equivalently 
\begin{equation}
\partial_t p_0(t;m) = -mz \tilde{w}_0p_0(t;m) + \tilde{w}p_1(t;m). \label{dp0dt}
\end{equation}
By applying the same reasoning to other levels, we find
\begin{eqnarray}
\partial_tp_1(t;m) &=& - (1+c_m)\tilde{w}p_1(t;m) + mz\tilde{w}_0p_0(t;m) + \tilde{w}p_2(t;m), \label{dp1dt} \\ 
\partial_t p_{n>1}(t;m) &=& - (1+c_m)\tilde{w}p_n(t;m) + c_m\tilde{w}p_{n-1}(t;m) + \tilde{w}p_{n+1}(t;m). \label{dpndt}
\end{eqnarray}
Here we have introduced the quantity $c_m = (mz-1)q$, with $q = \int^{\mathcal{C}k_BT}_{-\infty} d(\Delta E) P(\Delta E)$ giving the likelihood of an edge being energetically favorable. $c_m$ gives the number of children of a node at level 1 or beyond. One can readily show that $\partial_t \sum_np_n(t;m) = 0$, i.e. the total probability ($=1$) is conserved. The equations imply slightly different properties of the root node, which is however an artifact that arises because of all the transition rates only that (i.e. $\tilde{w}_0$) from level 0 to 1 can be computed exactly while the rest ($\tilde{w}$) are approximations. 

Equation (\ref{w}) provides $\tilde{w}$ by averaging the transition time $w(\Delta E)^{-1}$. 
At low temperatures, the long-time dynamics is dominated by a series of transitions along the levels on a thin tree branch with a degree $c_m$ close to unity for some dominant value of $m$. The average of the transition time, as opposed to that of the transition rate, thus gives a more accurate description for long-time dynamics.
Furthermore, Eq. (\ref{w}) is accurate even at higher temperatures. This is because $w(\Delta E)$ then admits small fluctuations and averaging over time or rate indeed give similar values.

From Eqs.~(\ref{dp0dt}) - (\ref{dpndt}), it follows that the average level 
\begin{equation}
\bar{n}(t;m) = \sum_n np_n(t;m), \label{nbar_sum} 
\end{equation}
the local region could have reached must satisfy the following equation,
\begin{equation}
\partial_t\bar{n}(t;m) = (c_m -1)\tilde{w} \left[1-p_0(t;m)\right] + mz\tilde{w}_0p_0(t;m), \label{dnbardt}
\end{equation}
subjected to initial conditions $\bar{n}(0;m) = 0$ and $p_0(0;m) = 1$. An integration gives
\begin{equation}
\bar{n}(t;m) = \int_0^t dt'\{ (c_m -1)\tilde{w} \left[1-p_0(t';m)\right] + mz\tilde{w}_0p_0(t';m) \}. \label{nbar} \end{equation}
It is interesting to see that $\bar{n}(t;m)$ is completely determined if $p_0(t;m)$ is known. For small $t$, we may take $p_0(t;m) \rightarrow 1$, and then $\bar{n}(t;m) \rightarrow mz\tilde{w}_0 t$, i.e. the initial growth of $\bar{n}(t;m)$ which is exact.  

Inspecting Eq.~(\ref{dnbardt}) or (\ref{nbar}), we note that there are two types of solutions depending on whether $c_m$ is greater or less than $1$. For $c_m>1$, $\bar{n}$ increases without bound, implying that the local region will be descending down the tree and access all of its configurations in time. In the meantime, $p_0(t;m)$ must decay to zero, yielding
\begin{equation}
	\label{nbar_large_t}
	\bar{n}(t;m) \approx (c_m-1)\tilde{w}t
\end{equation}
for large $t$. We may call such regions \textit{mobile regions}. 

For $c_m<1$, however, $\bar{n}$ cannot increase arbitrarily. Instead, it will initially increase as the second term on the right-hand side of Eq.~(\ref{dnbardt}) is larger than the first term, but finally saturate when the two terms balance each other. In this regime, the local region is trapped in a few configurations, and cannot escape. These regions may be called \textit{immobile regions}.
 
\subsection{Solutions to the rate equations}
\label{sec:Solutions_to_the_rate_equations}
While Eqs.~(\ref{dp0dt}) - (\ref{dpndt}) may be numerically integrated in a straightforward manner, here we solve them by the method of Laplace transform. So we define
\begin{equation}
	\label{Pn_def}
	P_n(s;m) = \int^\infty_0 dt ~ e^{-st} p_n(t;m),
\end{equation}
where $s$ resides in the upper half complex plane. In terms of $P_n(s;m)$, Eqs.~(\ref{dp0dt}) and (\ref{dp1dt}) can be written as
\begin{eqnarray}
&~& sP_0(s;m) - 1 = -mz\tilde{w}_0P_0(s;m) + \tilde{w}P_1(s;m), \label{sP0} \\
&~& sP_1(s;m) = -(1+c_m)\tilde{w}P_1(s;m) + mz\tilde{w}_0P_0(s;m) + \tilde{w}P_2(s;m). \label{sP1}
\end{eqnarray}
Here we have used the initial condition that $p_n(0;m) = \delta_{n,0}$. Expressing $P_{n=1,0}$ in terms of $P_2$, we find
\begin{equation}
P_{n}(s;m) = \alpha_{n}(s;m) + \beta_{n}(s;m)P_2(s;m), \label{P01}
\end{equation}
where $\alpha_{n=0,1}$ and $\beta_{n=0,1}$ are some coefficients given as follows
\begin{eqnarray}
	\label{alpha_beta}
	\squarebk{
		\begin{tabular}{ccc}
			$\alpha_{0}(s;m)$ & $\beta_{0}(s;m)$ \\
			$\alpha_{1}(s;m)$ & $\beta_{1}(s;m)$
		\end{tabular}
	}
	&=& \frac{1}{(s + mz\tilde{w}_{0})\squarebk{s + (1+c_{m})\tilde{w}} - mz\tilde{w}_{0}\tilde{w}} \nonumber \\
	& & \times \squarebk{
		\begin{tabular}{ccc}
			$s + (1 + c_{m})\tilde{w}$ & $\tilde{w}^{2}$ \\
			$mz\tilde{w}_{0}$ & $\tilde{w}(s + mz\tilde{w}_{0})$
		\end{tabular}	
	}.
\end{eqnarray}
Analogously, Eq.~(\ref{dpndt}) is transformed as
\begin{equation}
sP_n(s;m) = c_m\tilde{w}\left[P_{n-1}(s;m)-P_n(s;m)\right] + \tilde{w}\left[P_{n+1}(s;m)-P_n(s;m)\right]. \label{sPn}
\end{equation}
This equation is invariant under a translation of $n$ and thus the general solution is of the form 
\begin{equation}
P_n(s;m) = P_2(s;m)c^{n-2}, \quad n>1. \label{Pn}
\end{equation}
The quantity $c$ is obtained by substituting this expression into (\ref{sPn}). We find
\begin{equation}
	\label{c_eqt}
	c^2-(1+c_m+s\tilde{w}^{-1})c + c_m = 0.
\end{equation}
This has two solutions. One solution leads to $P_n$ that diverges for large $n$ and must be excluded. The other solution is given by 
\begin{equation}
c(s;m) = \frac{1}{2}\left(1+c_m+s\tilde{w}^{-1} - \sqrt{(1+c_m+s\tilde{w}^{-1})^2 - 4c_m}\right).
\label{c}  
\end{equation}
Finally, the normalization condition of the probabilities implies that
\begin{equation}
\sum^\infty_{n=0} P_n(s;m) = \int^\infty_0 dt e^{-st} = s^{-1}. 
\end{equation}
Combining this with Eqs.~(\ref{P01}) and (\ref{Pn}), we obtain
\begin{equation}
\label{P2}
P_2(s;m) = \frac{s^{-1} - \alpha_0(s;m) - \alpha_1(s;m)}{\left[1-c(s;m)\right]^{-1}+\beta_0(s;m) + \beta_1(s;m)},
\end{equation}
which can be used to generate all $P_n(s;m)$. Going back to $p_n(t;m)$, we need to invert the Laplace transform \begin{equation}
	\label{inverse_LP}
	p_n(t;m) = \frac{1}{2\pi i}\int^{\gamma+i\infty}_{\gamma -i\infty} ds~e^{st} P_n(s;m),
\end{equation}
which can be numerically evaluated using standard algorithms~\cite{valsa1998,dingfelder2015}. 

\section{Calculating \textbf{MSD}$(t)$}
\label{sec:Calculating_MSD}
The migration of a region on the configuration tree can be mapped onto the real-space motions of the particles in this region, thereby allowing us to calculate the MSD, \textbf{MSD}$(t)$ of these particles. 

To this end, let us observe that, as the region moves down the tree by one level, what happens in real space is that a void propagates along a string and each particle in this string hops by an average distance $a$ comparable to the atomic separation.
Suppose the region has descended to level $n$ in a period $t$.
During this period, in total $n\tilde{l}$ particle hops have occurred in $n$ strings each of length $\tilde{l}$, excluding the hops due to back-and-forth repetitions between any pair of nodes, which are highly correlated and do not contribute to the MSD.
Some particles may have hopped more than once.
As long as the hops are uncorrelated, as is statistically reasonable for \textit{non-repetitive} hops, the squared displacement of a particle over ensemble average is simply $a^2$ times the number $n_{hop,j}$ of non-repetitive hops particle $j$ has executed, i.e. $\langle\mathbf{d}^2_j\rangle = n_{hop,j}a^2$ with $\sum_jn_{hop,j} = n\tilde{l}$.
The contribution of each non-repetitive hop to the MSD is just $a^2/\mathcal{V}\phi$, where $\phi$ is the particle density and $\mathcal{V}\phi$ is the number of particles in the region.
As such, for a region reaching level $n$, the MSD is $n\tilde{l}a^2/\mathcal{V}\phi$.
Ensemble averaging this by $p_n(t;m)$, we obtain the MSD for this region as
\begin{equation}
	\label{MSD_eqt}
	\mbox{\textbf{MSD}}(t;m) = \frac{\tilde{l}a^2}{\mathcal{V}\phi} \bar{n}(t;m).
\end{equation}
Here $\bar{n}(t;m)$ can be directly obtained from Eq.~(\ref{nbar}) with $p_0(t;m)$ provided in Sec.~\ref{sec:Solutions_to_the_rate_equations}. 

To obtain the MSD for the entire liquid, we have to take an average of \textbf{MSD}$(t;m)$ over all local regions. The number of voids in one region may differ from that in another. Suppose the average number of voids per region is $\bar{m} = \mathcal{V}\phi_v$, where $\phi_v$ is the density of voids. With small correlations between voids, their steady-state distribution over the regions then approximately follows Poisson's law, i.e. the probability of a region with $m$ voids is 
\begin{equation}
	\label{fm}
	f(m;\bar{m}) = \frac{\bar{m}^me^{-\bar{m}}}{m!}.
\end{equation}
Now averaging \textbf{MSD}$(t;m)$ by this distribution, we finally arrive at
\begin{equation}
\mbox{\textbf{MSD}}(t) = \sum_m f(m;\bar{m})\mbox{\textbf{MSD}}(t;m) =  \frac{\tilde{l}a^2}{\mathcal{V}(\Omega^{-1}-\phi_v)} \sum_m f(m;\bar{m})\bar{n}(t;m).\label{MSD}
\end{equation} 
Here $\Omega$ is the volume of a particle or of a void, i.e. $\Omega^{-1} = \phi+\phi_v$. Note that Eq.~(\ref{MSD}) is closed with Eq.~(\ref{nbar}), which gives $\bar{n}(t;m)$. 

From Eq.~(\ref{MSD}), we can obtain the diffusion coefficient $D$ for the liquid. By definition, 
\begin{equation}
	\label{D_def}
	D = \lim_{t\rightarrow\infty} \frac{\mbox{\textbf{MSD}}(t)}{2d t},
\end{equation}
where $d$ is the dimension of the system. As discussed in Sec.~\ref{sec:Dynamics_on_the_tree_rate_equations}, for mobile regions with $c_m = (mz-1)q>1$, $\bar{n}(t;m) = (c_m-1)\tilde{w}t$ for $t\rightarrow\infty$. For immobile regions, however, $\bar{n}(t;m)$ approaches a constant for $t\rightarrow\infty$. Thus, only mobile regions contribute to $D$. We find
\begin{equation}
D =  \frac{\tilde{w}\tilde{l}a^2\Omega}{2d\mathcal{V}(1-\phi_v\Omega)} {\sum_m}' f(m;\bar{m})(c_m-1), \label{D}
\end{equation} 
where the prime restricts the summation to mobile regions. This result on $D$ was first obtained in Ref.~\cite{lam2018}, where the behavior of $D$ has been thoroughly investigated. 

\section{Comparison with simulations}
\label{sec:Comparison_with_simulations}
While the theory is expected to be valid for a variety of glassy systems exhibiting string-like motions, here we check it with the DPLM, for which most of the required parameters are known precisely and adjustable parameters are few. 
As it is a lattice model without vibrations, our theory can be directly applied. 
For comparison with MD simulations or experiments in the presence of vibrations, as aforementioned, the MSD must be based on highly coarse-grained trajectories with vibrations completely eliminated, such as those of inherent structures~\cite{liao2001,stillinger1982,stillinger1984,stillinger1995}.
While our focus is on the DPLM, we wish to point out that simulations have been done in Ref.~\cite{liao2001} by Liao \textit{et al} on the inherent structures of a binary Lennard-Jones liquid and their results are in good qualitative agreement with our theory.
A very important result of their work is that, the MSD of inherent structures initially increases linearly (rather than parabollically) in time and then goes through a shallow plateau before finally entering the diffusion regime.
These features are also borne out in our theory.  

As the DPLM has already been studied thoroughly in Ref.~\cite{zhang2017}, here a short description is provided for the convenience of readers. The DPLM assumes a number of distinguishable hard-core particles living on a square lattice with lattice constant $a=1$. Interactions occur between neighboring particles in the form $V_{ijs_is_j}$, each of which is sampled from a distribution $g(V)$ before a simulation begins. Here $i$ and $j$ denote a pair of neighboring sites while $s_i$ and $s_j$ denote the types of the particles sitting at the sites, respectively. Each particle is of its own type. A void is simply the absence of a particle. We have chosen $g(V)$ to be a uniform distribution in the range of $V \in [-0.5, 0.5]$. The string length is chosen to be $\tilde{l} = 1$. The kinetic Monte Carlo rules of activated hopping is employed for the dynamics of the particles. Each particle can hop to an empty neighboring site at a rate 
\begin{equation}
w(\Delta E) = w_0e^{-(E_0+\Delta E/2)/k_BT},
\label{wDE}
\end{equation}
where $w_0$ and $E_0$ are constants and $\Delta E$ is the change in the system energy due to the hop. Though the equilibrium properties of this model are exactly solvable, it displays many characteristic behaviors, including the two-step relaxation of a glass former~\cite{zhang2017}. 

Note that $\Delta E$ is also the energy change due to a void propagation. In the configuration-tree theory, we call an edge in the local tree energetically favorable if $\Delta E$ for the corresponding void propagation is less than an upper cutoff $\mathcal{C}k_BT$, where $\mathcal{C}$ is a parameter of the order of unity. Thus, the probability $q$ that an edge is energetically favorable can be directly evaluated from $g(V)$ and the equilibrium properties of the DPLM, see Ref.~\cite{zhang2017}. Since $\Omega=1$ for the DPLM, this parameter $\mathcal{C}$ and the volume of a local region $\mathcal{V}$ are the only parameters needed to evaluate \textbf{MSD}$(t)$ by Eq.~(\ref{MSD}). These parameters can actually be further fixed by fitting the analytical expression of $D$, Eq.~(\ref{D}) to the numerical simulations of the DPLM. This has been carried out in Ref.~\cite{lam2018} using $\tilde w=w_0$, where it was found that $\mathcal{C} = 1.7$ and $\mathcal{V}=12$ give the best fit over a wide range of temperatures and void densities. This choice is appropriate for both the activated hopping dynamics in (\ref{wDE}) and the Metropolis dynamics, with a wide choice of $g(V)$, where $\Delta E$ averages to 0. 
Details on the Metropolis dynamics will be given elsewhere. As such, virtually no free parameters are involved in the analytical calculation of \textbf{MSD}(t).

Using the Laplace-transformation approach, by applying  
Eq.~(\ref{P2}) together with the coefficients from Eqs.~(\ref{alpha_beta}) and (\ref{c}) as well as
the rates defined by Eq.~(\ref{w_0}) and Eq.~(\ref{w}), one can calculate $P_2(s;m)$. Then, $P_0(s;m)$ follows from Eqs.~(\ref{sP0}) and (\ref{sP1}) and can be inserted into Eq.~(\ref{inverse_LP}) to obtain $p_0(t;m)$.
Substituting $p_{0}(t;m)$ into Eq.~(\ref{nbar}), we find $\bar{n}(t;m)$, which is further inserted in Eq.~(\ref{MSD}) to obtain the MSD. Eq.~(\ref{MSD}) and these associated equations thus constitute our closed form solution for the MSD.
Alternatively, identical results can be obtained using an ordinary differential equation (ODE) integration approach. In this case, the rate equations in Eqs.~(\ref{dp0dt}) - (\ref{dpndt}) are numerically integrated to obtain, in particular, $p_0(t;m)$. The MSD can then be obtained similarly using Eqs.~(\ref{nbar}) and (\ref{MSD}).

\begin{figure}[tb]
	\label{f:msd}
	\includegraphics[width=0.5\linewidth]{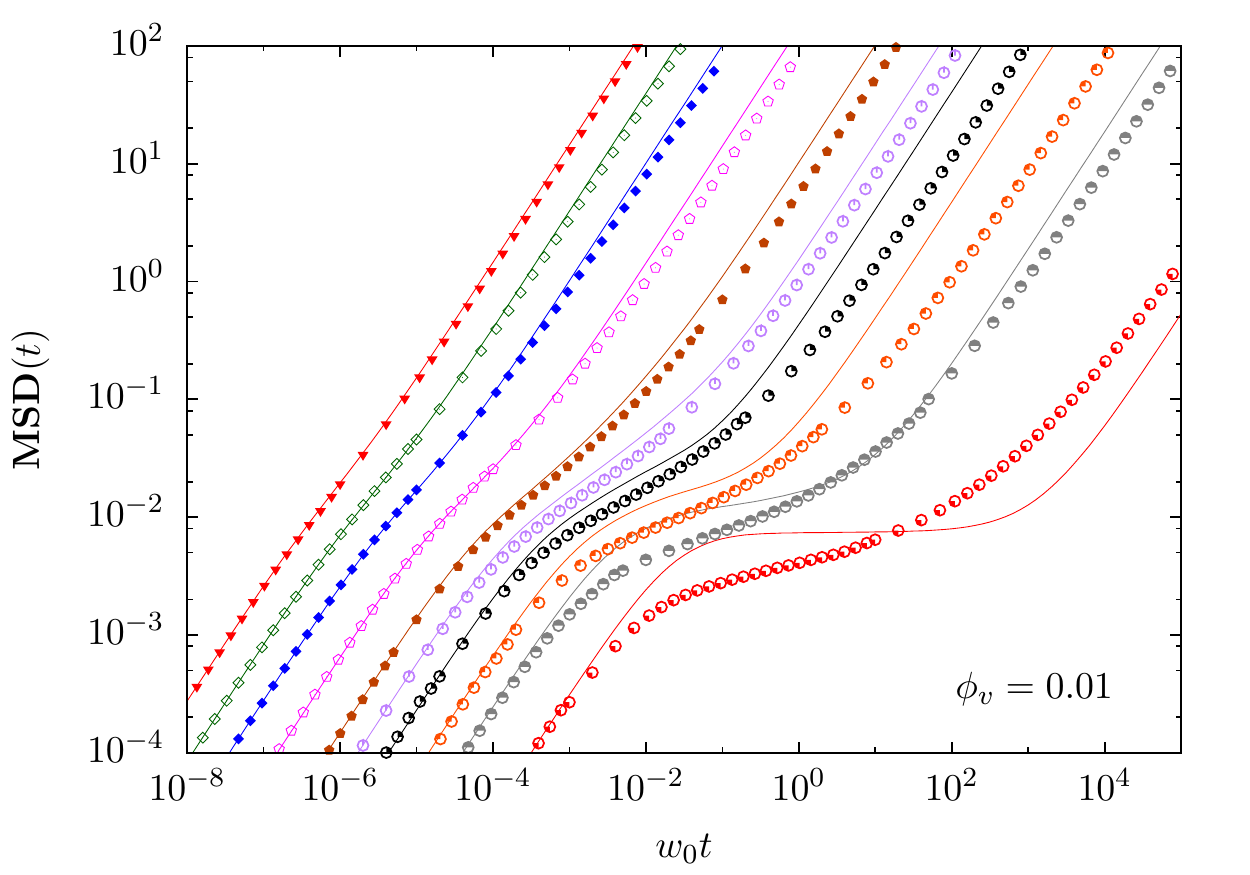}
	\includegraphics[width=0.5\linewidth]{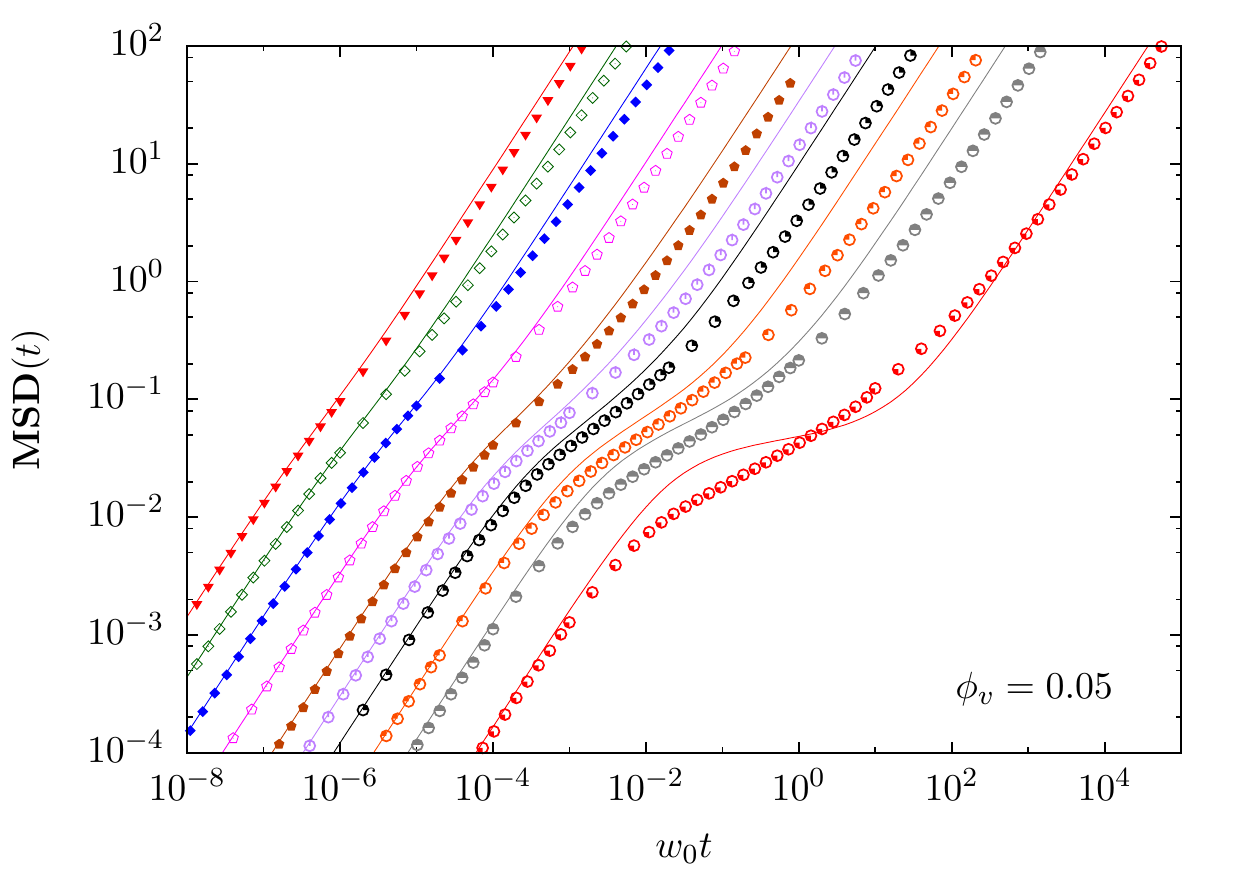}
	\caption{Comparison of the mean-squared displacement \textbf{MSD}(t) versus time $t$ between the theory (lines) and the DPLM simulation (points), with void densities $\phiv$ = $0.01$ (left panel) and $\phiv$ = $0.05$ (right panel). Different curves represent systems at different temperatures, $T = 4, 1, 0.6, 0.4, 0.3, 0.26, 0.23, 0.2, 0.18, 0.15$ (from left to right).}
\end{figure}

The agreement of the theory with the simulations for a wide range of temperatures is shown in Figure~2. In the plot, we have chosen a void density $\phi_v = 0.01$ for the left panel while $0.05$ for the right panel. 
The temperature values range from the non-glassy liquid phase at $T=4$ to the deeply supercooled phase at $T=0.15$ \cite{zhang2017}.  Here,  glassy behaviors at low $T$ manifest as a plateau, which is relatively shallow only because of the lack of vibrations in the model and by no means implying a slight or moderate supercooling. Indeed, this shallow plateau is also a property of the MSD observed in lattice models \cite{kob1993} and of inherent structures of a binary Lennard-Jones liquid~\cite{liao2001}. As shown in the figure, the agreement is very good at high temperatures, while at low temperatures the theory deviates from the simulations but the agreement is still impressive. The nearly exact agreement for \textbf{MSD}$(t) \leq 2\times 10^{-3}$ for all curves also shows the validity of the rates $\tilde{w}_0$ used in our theory. More importantly, the theory captures the location of the plateau very well for the full wide range of temperatures studied.
Only at very low temperatures $T \le 0.2$, the predicted plateaus become too flat compared with the more shallow ones from simulations. The 
discrepancy may be due to neglecting random fluctuations in the theory as it adopts averaged transition rates and assumes local regions completely isolated from other parts of the system.
At any finite $T$, our theory predicts no ideal phase transition \cite{lam2018}.
Therefore, even at arbitrarily low but finite $T$, an increasingly broad plateau in the MSD is predicted, which eventually must cross over to the long-time diffusive regime.

\section{Conclusion and outlook}
\label{sec:Conclusion_and_outlook}
Based on the local random configuration-tree theory, we have analytically calculated the MSD based on the inherent structures of glass formers. The results have been compared with numerical simulations of a lattice glass model -- the DPLM, and an overall satisfactory agreement has been achieved at the quantitative level with few free parameters.

In this paper we have only applied the theory to the DPLM, for the required parameters are precisely known here. In the near future, we shall apply it to other glass formers, including computational (e.g. polydispersed Lennard-Jones liquids) and experimental systems. For such systems, the MSD should be based on coarse-grained trajectories~\cite{lam2017} for inherent structures with fast vibrations averaged out. There will be far more parameters involved. However, these parameters fall in groups and each group can be tightly constrained by computational or experimental observations. We shall also use the theory to shed light on the effects of particle swap on particle dynamics in these systems, which has recently attracted lots of attention~\cite{berthier2016,ninarello2017,ikeda2017,brito2018,szamel2018,wyart2017,berthier2017}. 

Another interesting line of investigation would be to look at other physical quantities from the configuration-tree theory point of view, such as the self-intermediate scattering function and the particle displacement distribution function. The former is especially important for its widespread use in the experimental literature. One should note that the MSD is closely related to this quantity in the long wavelength limit. We are currently working on generalizing our theory to calculate these functions. 

Finally, we make a few remarks on the role of voids in our theory. There are two basic tenets in the theory. The first is the existence of a local random configuration tree, which in our opinion is quite general and robust. The second explicitly assumes the presence of voids and string-like particle motions. While string-like motions have ample evidences, the concept of voids is sometimes contested. Regarding this the following point is worthy of notice: A void capable of initiating string-like motions is not necessarily as well-defined spatially as in lattice gas models such as the DPLM. It could be a quasi-void consisting of a relatively small void coupled to free volumes fragmentary and distributed  \cite{lam2017}.
\ack

We thank Giorgio Parisi for helpful comments. We are grateful for the support of Hong Kong GRF (Grant 15330516) and Hong Kong PolyU (Grant 1-ZVGH).

\

\section*{Appendix: Summary of main assumptions and comparison with the literature}

Here we summarize the main concepts and implications of the theory and discuss their similarities and differences to other approaches in the literature.

\begin{enumerate}
\item {\bf The starting assumption of the tree theory is that elementary particle motions are activated particle hops initiated by voids}, an idea that can be traced back to Glarum's theory of diffusion of defects identifiable as units of free volume or voids \cite{glarum1960} and activated motions introduced by Goldstein \cite{goldstein1969}. A hop takes the system from one inherent structure \cite{stillinger1983} to another. The hopping distance is assumed fixed and comparable to the particle diameters, as measured from the short-time van Hove correlation function \cite{wahnstrom1991,marcus1999,lam2017}. 
Our theory is thus microscopic, based on elementary atomistic processes, unlike for example continuous-time random walk (CTRW) models \cite{montroll1965,chaudhuri2007,helfferich2014}, which assumes a broad distribution of hopping distances appropriate at a more coarse-grained level. Voids of volumes comparable to particle sizes may be too few to account for observed motions \cite{starr2002}. Noting the non-rigidity of fluids, in our theory we conceive that the a void may also be a quasi-void which consists of a relatively small interstitial void and fragmented free volumes close by, so that the squeezing-in of another particle can happen \cite{lam2017}.

In addition, particle hops can be correlated in the form of microstrings \cite{glotzer2004,chandler2011} each describing simultaneous coherent hops of a line of particles. The length of microstrings is short and narrowly distributed \cite{glotzer2004} and thus approximately fixed. This is in contrast with the length of incoherent strings, which has been phenomenologically assumed to diverge upon cooling\cite{langer2006}.

\item {\bf The second assumption is that the particles are assumed distinguishable with particle-dependent pair interactions.} Particles with diverse properties have been modeled for example in Refs.~\cite{pedersen2006,rabin2015}. This is directly justifiable for polydispersive and polymer systems. It may also be applicable to identical particles as it effectively models different frustration states at various meta-stable particle positions \cite{zhang2017}. 
In contrast, however, most studies including defect models usually assume particles with identical properties \cite{glarum1960,fredrickson1984,kob1993,biroli2001,langer2006}.

\item {\bf As a consequence of (ii), the system energy change involved in a particle hop then follows from the system potential energy landscape (PEL), which is a function defined in the particle configuration space of distinguishable particles.} 
Unlike spin-glass models, the PEL accounts for quenched disorder in the configuration space but not in the real space \cite{edwards1975}.
The configuration space contains all possible inherent structures of the system. This space is substantially larger than that of identical particles because the exchange of particles, for example, leads to a different configuration with in general a different energy. The PEL is a rugged energy landscape \cite{goldstein1969,bassler1987}. Hops of particles and voids are thus history dependent, as further explained in point (vi), in contrast to history independent hops in many theories \cite{glarum1960,langer2006,phan2018}.
Motions of voids are then described by random walks in the configuration space, analogous to previous  descriptions for particles in Refs. \cite{bassler1987,dyre1995}. 

\item {\bf A further consequence of (ii) is that the system energy change depends on the paths of the voids and in particular on the time-order of the paths.} We have described the phenomenon as "path interactions of voids'' or equivalently "string interactions" \cite{lam2017}. 
This is because a void's motions alter the particle configurations along its whole track and perturb subsequent motions of other voids. 
Therefore, the hop of a void changes the hopping rate of another void close by.
In contrast, voids are assumed non-interacting usually \cite{glarum1960}.

\item {\bf In accord with (iii), the particle configuration space for a local region with $m$ voids is approximated by the Bethe lattice with a coordination number proportional to $m$. Each edge represents a hop associated with an independent energy change.} The configuration space is of a very high dimension and is much closer to a tree than the physical space. For a local region with a single void, the tree geometry is indeed exact up to the fifth tree level for our lattice model \cite{zhang2017}. 
For a region with multiple voids, a fully interacting void approximation also leads to a tree geometry with independent edge properties \cite{lam2018}. 
Discrepancies result effectively at a reduced number of independent tree branches. 
Nevertheless, qualitative features of our theory are robust as long as such an effective tree coordination number increases with $m$.

\item {\bf As an approximation, the nodes in the Bethe lattice are classified as energetically favorable or unfavorable. The favorable nodes constitute a random tree that is assumed to have a flat energy landscape.} Within the random tree which is a sub-tree of the Bethe lattice, our theory reduces to a purely kinetic one analogous to the kinetically constrained models (KCM) \cite{garrahan2011review}.
A key feature of our theory is that the energetics, i.e. the PEL with the disorder quenched in the configuration space, is accounted for by the geometry of the random tree. The quenched disorder and the resulting history dependence of voids motions are then treated as the confinement to a fixed sub-tree in the configuration space. It is closely analogous to, for example, studying diffusion in fractal channels constituting quenched disorder in the physical space in porous media \cite{Sahimibook}.

\item {\bf Particle mobility in a local region with $m$ voids reduces to an $m$-dependent percolation problem of the random tree in the Bethe lattice.  }
Since the underlining Bethe lattice has a coordination number increasing with $m$, the degree (average number of children per node) of the random tree increases with $m$. The  void mobility and hence the particle mobility also increase with $m$, realizing a facilitation phenomenon. Percolation theory has been applied in the configuration space \cite{dyre1995}, but is more often applied in the physical space \cite{cohen1979}.

\item {\bf A higher mobility at regions with multiple voids is a facilitation phenomenon.}  Physically,  at a temperature at which a single void is typically described by a finite tree, the void is trapped among a few meta-stable sites. With another void, its passing by perturbs the particle configuration so that the first void may  have access to additional configurations, accounted for by additional tree-branches. This  can untrap the first void temporarily. Such mutual facilitation of motions between the voids can lead to a mobile pair of interacting voids. This is analogous to the two-spin facilitation model of Fredrickson and Andersen \cite{fredrickson1984}, one of the first KCM \cite{garrahan2011review}. Our theory provides a physical explanation of the facilitation rules.
In contrast, trap models \cite{odagaki1990} consider untraping via surmounting energy barriers by visiting also the energetically unfavorable configurations. They offer however slower untrapping dynamics negligible compared with the facilitation mechanism.

\item {\bf Super-Arrhenius dynamics and kinetic arrest follow from a sequence of mobility transition of increasing $m$ upon cooling.} As $T$ decreases, regions with one, two, three voids and so on successively enter the immobile phase resulting at dramatic slowdown. Our theory is thus analogous to the  $p$-spin facilitation model \cite{fredrickson1984}, with $p$ increasing upon cooling. 

\end{enumerate}

%%% 

\

\

\noindent\textbf{References}

\

\bibliographystyle{iopart-num}
\bibliography{polymer_short}

% Our own papers related to the tree-theory
  %%%%%%%%%%%%%%%%%%%%%%%%%%%%%%%%%%%%%%%%%%%%%%% % Polymer film PRE
  %%%%%%%%%%%%%%%%%%%%%%%%%%%%%%%%%%%%%%%%%%%%%%% % String repetition
  %%%%%%%%%%%%%%%%%%%%%%%%%%%%%%%%%%%%%%%%%%%%%%%%%%%% % UGC15
  %%%%%%%%%%%%%%%%%%%%%%%%%%%%%%%%%%%%%%%%%%%%%%%%%%%% % DPLM
  %%%%%%%%%%%%%%%%%%%%%%%%%%%%%%%%%%%%%%%%%%%%%%%%%%%% % tree-theory
  %%%%%%%%%%%%%%%%%%%%%%%%%%%%%%%%%%%%%%%%%%%%%%%%%%%% % ugc17
  %%%%%%%%%%%%%%%%%%%%%%%%%%%%%%%%%%%%%%%%%%%%%%%%% % polymer surface
  %%%%%%%%%%%%%%%%%%%%%%%%%%%%%%%%%%%%%%%%%%%%%%%%%%%% % dplm-notes
  %%%%%%%%%%%%%%%%%%%%%%%%%%%%%%%%%%%%%%%%%%%%%%%%%%%% % UGC18
  %%%%%%%%%%%%%%%%%%%%%%%%%%%%%%%%%%%%%%%%%%%%%%%%%%%% % MSD
  %%%%%%%%%%%%%%%%%%%%%%%%%%%%%%%%%%%%%%%%%%%%%%%%%%%%
\providecommand{\newblock}{}
\begin{thebibliography}{10}
\expandafter\ifx\csname url\endcsname\relax
  \def\url#1{{\tt #1}}\fi
\expandafter\ifx\csname urlprefix\endcsname\relax\def\urlprefix{URL }\fi
\providecommand{\eprint}[2][]{\url{#2}}
% Bibliography created with iopart-num v2.1
% /biblio/bibtex/contrib/iopart-num

\bibitem{berthier2011review}
Berthier L and Biroli G 2011 {\em Rev. Mod. Phys.\/} {\bf 83} 587

\bibitem{garrahan2011review}
Garrahan J~P, Sollich P and Toninelli C {\em in Dynamical Heterogeneities in
  Glasses, Colloids and Granular Media, edited by L. Berthier, G. Biroli, J.-P.
  Bouchaud, L. Cipelletti, and W. van Saarloosand (Oxford University Press,
  2011)\/}

\bibitem{stillinger2013review}
Stillinger F~H and Debenedetti P~G 2013 {\em Annu. Rev. Condens. Matter
  Phys.\/} {\bf 4} 263

\bibitem{angell2000jap}
Angell C~A, Ngai K~L, McKenna G~B, McMillan P~F and Martin S~W 2000 {\em J.
  Appl. Phys.\/} {\bf 88} 3113

\bibitem{Gotzebook}
G{\H o}tze W 2008 {\em {Complex dynamics of glass-forming liquids: a
  mode-coupling theory}\/} (Oxford University Press)

\bibitem{lam2018}
Lam C~H 2018 {\em J. Stat. Mech.\/} {\bf 2018} 023301

\bibitem{fredrickson1984}
Fredrickson G~H and Andersen H~C 1984 {\em Phys. Rev. Lett.\/} {\bf 53} 1244

\bibitem{palmer1984}
Palmer R~G, Stein D~L, Abrahams E and Anderson P~W 1984 {\em Phys. Rev.
  Lett.\/} {\bf 53} 958

\bibitem{ritort2003review}
Ritort F and Sollich P 2003 {\em Adv. Phys.\/} {\bf 52} 219

\bibitem{chandler2011}
Keys A~S, Hedges L~O, Garrahan J~P, Glotzer S~C and Chandler D 2011 {\em Phys.
  Rev. X\/} {\bf 1} 021013

\bibitem{isobe2016}
Isobe M, Keys A~S, Chandler D and Garrahan J~P 2016 {\em Phys. Rev. Lett.\/}
  {\bf 117}(14) 145701

\bibitem{mandadapu2018}
Katira S, Garrahan J~P and Mandadapu K~K 2018 {\em Phys. Rev. Lett.\/} {\bf
  120} 260602

\bibitem{helfferich2014}
Helfferich J, Ziebert F, Frey S, Meyer H, Farago J, Blumen A and Baschnagel J
  2014 {\em Phys. Rev. E\/} {\bf 89} 042603

\bibitem{phan2018}
Phan A~D and Schweizer K~S 2018 {\em J. Phys. Chem. B\/} {\bf 122} 8451

\bibitem{zhang2017}
Zhang L~H and Lam C~H 2017 {\em Phys. Rev. B\/} {\bf 95}(18) 184202

\bibitem{lam2018film}
Lam C~H 2018 {\em J. Chem. Phys.\/} {\bf 149} 164909

\bibitem{lam2017}
Lam C~H 2017 {\em J. Chem. Phys.\/} {\bf 146} 244906

\bibitem{glotzer2004}
Gebremichael Y, Vogel M and Glotzer S~C 2004 {\em J. Chem. Phys.\/} {\bf 120}
  4415

\bibitem{stillinger1983}
Stillinger F~H and Weber T~A 1983 {\em Phys. Rev. A\/} {\bf 28} 2408

\bibitem{liao2001}
Liao C~Y and Chen S~H 2001 {\em Phys. Rev. E\/} {\bf 64} 031202

\bibitem{cassi1989}
Cassi D 1989 {\em Euro. Phys. Lett.\/} {\bf 9} 627

\bibitem{hughes1982}
Hughes B~D and Sahimi M 1982 {\em J. Stat. Phys.\/} {\bf 29} 781

\bibitem{valsa1998}
Valsa J and Brancik L 1998 {\em Int. J. Numer. Model.\/} {\bf 11} 153

\bibitem{dingfelder2015}
Dingfelder B and Weideman J~A~C 2015 {\em Numer. Algor.\/} {\bf 68} 167

\bibitem{stillinger1982}
Stillinger F~H and Weber T~A 1982 {\em Phys. Rev. A\/} {\bf 25}(2) 978

\bibitem{stillinger1984}
Stillinger F~H and Weber T~A 1984 {\em Science\/} {\bf 225} 983

\bibitem{stillinger1995}
Stillinger F~H 1995 {\em Science\/} {\bf 267} 1935

\bibitem{kob1993}
Kob W and Andersen H~C 1993 {\em Phys. Rev. E\/} {\bf 48}(6) 4364

\bibitem{berthier2016}
Berthier L, Coslovich D, Ninarello A and Ozawa M 2016 {\em Phys. Rev. Lett.\/}
  {\bf 116} 238002

\bibitem{ninarello2017}
Ninarello A, Berthier L and Coslovich D 2017 {\em Phys. Rev. X\/} {\bf 7}
  021039

\bibitem{ikeda2017}
Ikeda H, Zamponi F and Ikeda A 2017 {\em J. Chem. Phys.\/} {\bf 147} 234506

\bibitem{brito2018}
Brito C, Lerner E and Wyart M 2018 {\em Phys. Rev. X\/} {\bf 8} 031050

\bibitem{szamel2018}
Szamel G 2018 {\em arXiv:1805.02753\/}

\bibitem{wyart2017}
Wyart M and Cates M~E 2017 {\em Phys. Rev. Lett.\/} {\bf 119} 195501

\bibitem{berthier2017}
Berthier L, Charbonneau P, Flenner E and Zamponi F 2017 {\em Phys. Rev.
  Lett.\/} {\bf 119} 188002

\bibitem{glarum1960}
Glarum S~H 1960 {\em J. Chem. Phys.\/} {\bf 33} 639

\bibitem{goldstein1969}
Goldstein M 1969 {\em J. Chem. Phys.\/} {\bf 51} 3728

\bibitem{wahnstrom1991}
Wahnstr{\"o}m G 1991 {\em Phys. Rev. A\/} {\bf 44} 3752

\bibitem{marcus1999}
Marcus A~H, Schofield J and Rice S~A 1999 {\em Phys. Rev. E\/} {\bf 60} 5725

\bibitem{montroll1965}
Montroll E~W and Weiss G~H 1965 {\em New York\/} {\bf 6} 167

\bibitem{chaudhuri2007}
Chaudhuri P, Berthier L and Kob W 2007 {\em Phys. Rev. Lett.\/} {\bf 99} 060604

\bibitem{starr2002}
Starr F~W, Sastry S, Douglas J~F and Glotzer S~C 2002 {\em Phys. Rev. Lett.\/}
  {\bf 89} 125501

\bibitem{langer2006}
Langer J~S 2006 {\em Phys. Rev. Lett.\/} {\bf 97} 115704

\bibitem{pedersen2006}
Pedersen U~R, Hecksher T, Dyre J~C and Schr{\o}der T~B 2006 {\em J. Non-Cryst.
  Solids\/} {\bf 352} 5210

\bibitem{rabin2015}
Shagolsem L~S, Osmanovi{\'c} D, Peleg O and Rabin Y 2015 {\em J. Chem. Phys.\/}
  {\bf 142} 051104

\bibitem{biroli2001}
Biroli G and M\'ezard M 2001 {\em Phys. Rev. Lett.\/} {\bf 88}(2) 025501

\bibitem{edwards1975}
Edwards S~F and Anderson P~W 1975 {\em Journal of Physics F: Metal Physics\/}
  {\bf 5} 965

\bibitem{bassler1987}
B{\"a}ssler H 1987 {\em Phys. Rev. Lett.\/} {\bf 58} 767

\bibitem{dyre1995}
Dyre J~C 1995 {\em Phys. Rev. B\/} {\bf 51} 12276

\bibitem{Sahimibook}
Sahimi M 2011 {\em Flow and transport in porous media and fractured rock: from
  classical methods to modern approaches\/} (John Wiley \& Sons)

\bibitem{cohen1979}
Cohen M~H and Grest G~S 1979 {\em Phys. Rev. B\/} {\bf 20} 1077

\bibitem{odagaki1990}
Odagaki T and Hiwatari Y 1990 {\em Phys. Rev. A\/} {\bf 41} 929

\end{thebibliography}

\end{document}